# Strong field, scale separated, ultra low viscosity dynamos

Andrey Sheyko[1], Christopher Finlay[2], Jean Favre[3] and Andrew Jackson[1]

[1]Earth and Planetary Magnetism Group, Institute of Geophysics, ETH Zurich.
[2]Division of Geomagnetism, DTU Space, Technical University of Denmark.
[3]Swiss National Supercomputing Centre, Lugano, Switzerland.
[*] Correspondence should be addressed to Andrey Sheyko; E-mail: andrei.sheiko@erdw.ethz.ch

**The mechanism by which the Earth's magnetic field is generated is thought to be thermal convection in the metallic liquid iron core. Here we present results of a suite of self-consistent spherical shell computations with ultra-low viscosities that replicate this mechanism, but for the first time using diffusivities of momentum and electric current that are vastly different. This leads to significant scale separation between magnetic and velocity fields, the latter being dominated by small scales. The leading force balance at large scales has a major role played by the Lorentz force whereas small scales require a different balance. In this dynamo dissipation is almost exclusively Ohmic, as in the Earth, with convection inside the so-called tangent cylinder playing a crucial role. and has significantly more magnetic energy than kinetic energy (as in the Earth). Our model suggests Ohmic dissipation in the Earth's core is as high as 10TW.**

**Single sentence summary:**

Numerical dynamos with different scales in velocity and magnetic fields present new dynamics and dissipation behaviour for Earth's core.

## Background

Earth's dynamo is generally considered to be driven by cooling of the core (radius $r = r_0$) and from latent heat and buoyancy associated with the crystallisation of the inner core (radius $r = r_i$). Complex motions **u** associated with this cooling mechanism act in concert with an existing magnetic field **B** to generate electrical currents by Faraday's Law, and these currents generate more magnetic field by Ampere's Law. This general picture of a magnetic field generator is termed a dynamo, but the actual details are more complex: the system is governed by the coupled momentum (Navier-Stokes), induction (pre-Maxwell) and energy equations that must be simultaneously satisfied. Of primary importance in the momentum equation is the presence of the Coriolis force, associated with the rapid rotation of the Earth; this effect is generally considered paramount in leading to a roughly dipolar field whose axis is located close to the rotation axis of the Earth.

Numerical solutions of these sets of equations (*1, 2*) have borne great fruit, but numerical limitations limit the reality of the computations performed thus far. Here we report on solutions closer to the geophysical regime than previously reported, in which the Coriolis and Lorentz forces are dominant, viscosity plays a minor role, and we observe scale separation between magnetic and velocity fields. Using a length scale $L = r_o - r_i$, fluid kinematic viscosity $\nu$ and rotation rate $\Omega$, the Ekman number $E = \nu/(2\Omega L^2)$ for these simulations is as low as $E = 3 \times 10^{-7}$, rarely achieved in numerical studies. Table 1 gives details of the dynamo solutions computed, together with one purely hydrodynamic simulation (HYDRO0) that removes the presence of the magnetic field altogether (see also Supplementary material). A critical parameter is the magnetic Prandtl number $Pr_m = \nu/\eta$, where $\eta$ is the magnetic diffusivity. Liquid metals in general, including at high pressure, have very small values for $Pr_m$, namely $O(10^{-5} - 10^{-6})$. Our simulations include some that reduce this value below all others previously reported, namely to a value of 0.05 in the case of model S4. The aim of the reduction in $Pr_m$ is to ensure that almost all energy is dissipated Ohmically, as in the Earth, rather than viscously. Figure (1) compares our models with others in the literature and shows that we have achieved this, with model S4 dissipating 91% of its energy Ohmically (this fraction is commonly called $f_{\text{ohm}}$).

| Name | $E/10^{-6}$ | $Ra$ | $q = Pr_m$ | $R_m$ | $Re$ |
|---|---|---|---|---|---|
| S0 | 1.1834 | $Ra_0$ | 0.20 | 63 | 315 |
| HYDRO0 | 1.1834 | $Ra_0$ | | | 457 |
| S1 | 1.1834 | $5 \cdot Ra_0$ | 0.20 | 180 | 900 |
| S2 | 1.1834 | $30 \cdot Ra_0$ | 0.20 | 933 | 4665 |
| S4 | 0.2959 | $30 \cdot Ra_0$ | 0.05 | 274 | 5480 |

Table 1: Control parameters of the numerical simulations. $E$ is the Ekman number, $Ra$ is the Rayleigh number, $Pr_m$ is the magnetic Prandtl number. Output characteristics are the magnetic Reynolds number $R_m$ and the conventional Reynolds number $Re$. The Prandtl number is unity for all runs and S0 is as in (*3*) with $Ra_0 = 219.7$.

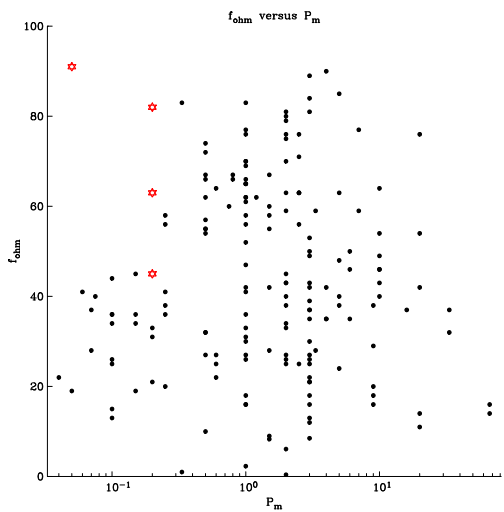

Figure 1: Fraction of dissipation that is Ohmic for both our models (red stars) and for the dataset from Uli Christensen based on models in (*4, 5*). The fraction is $f_{\text{ohm}} = D_{mag}/(D_{mag} + D_{kin})$ and is considered to be close to unity in planetary cores. Our models fill the low-$Pr_m$ parameter space while dissipating primarily Ohmically.

Our simulations in the electrically-conducting fluid outer and solid inner cores ($r_i/r_o = 0.35$) are constructed with the energy to drive the dynamos partitioned equally between heating at the inner core boundary (ICB; simulating crystallisation) and internal heating (secular cooling) (*6*). We use constant heat flux boundary conditions (*3*) at the core-mantle boundary (CMB) and a constant temperature ICB; no slip boundary conditions are applied at the ICB and CMB and the inner core is conducting and free to rotate. Details of our computations, which are standard, are given in the Supplementary materials. Kinetic Reynolds numbers in excess of 5000 are achieved in our most extreme calculation S4; see Table 1. The unit for magnetic field we use is the "Elsasser unit" $\sqrt{2\rho\Omega\mu_0\eta}$ where $\rho$ and $\mu_0$ are density and free space magnetic permeability respectively. With the recent values for $\eta \approx 0.5 m^2 s^{-1}$ (7), one Elsasser unit is close to 1mT. We have run the simulations, which are computationally demanding (requiring more than 90M CPU hours), until an equilibrium is found. All of our solutions are quasi-steady, dipole-dominated and non-reversing.

Fixed heat flux boundary conditions were suggested by Sakuraba and Roberts (3) as being more realistic boundary conditions for the core considering the overlying convecting mantle. They showed that the use of these boundary conditions affected the length scale of magnetic fields. Their simulation corresponds to our Case S0. Case S0 exhibits westward drift at the equator as a result of thermal winds driven by development of a hot equator. Convection has not set in within the tangent cylinder. When the Rayleigh number is raised by a factor of five, Case 1 exhibits convection within the tangent cylinder and reduced CMB temperature gradients, with concomitant reduced westward drift. As the Rayleigh number is raised to thirty times that of Case S0, the sign of the temperature gradient is reversed, leading to a cold equator, hot poles and a weak eastward drift. Case S4, which represents our most extreme calculation, has the same Rayleigh number as Case 2 but has had its Ekman and magnetic Prandtl numbers reduced by factors of four. Now (see Figure2) the tangent cylinder is very hot, but temperature gradients outside the TC are ameliorated to an extent that there is essentially no global azimuthal flow at the equator of the CMB. Thus the heat flux boundary condition plays a vital role in determining whether there is azimuthal flow at the equator or not.

Our simulations show evidence for the role of the Lorentz force in modifying the flow at all scales. Figure 3(a) compares non-magnetic and magnetic simulations at the same Rayleigh, Ekman and Prandtl numbers (Case S0 and Case HYDRO0). The magnetic field generates a Lorentz force that substantially modifies the energy content at all scales. Note that in the HYDRO case we have a very good fit to Kolmogorov's $-5/3$ law for the energy spectrum, despite the fact that this law is generally applied to isotropic non-rotating turbulence. Figure 3(b) shows our extreme simulation at $E = 3 \times 10^{-7}, Pr_m = 0.05$ (Case S4). Magnetic energy exceeds kinetic energy for all spherical harmonic degrees $l < 100$.

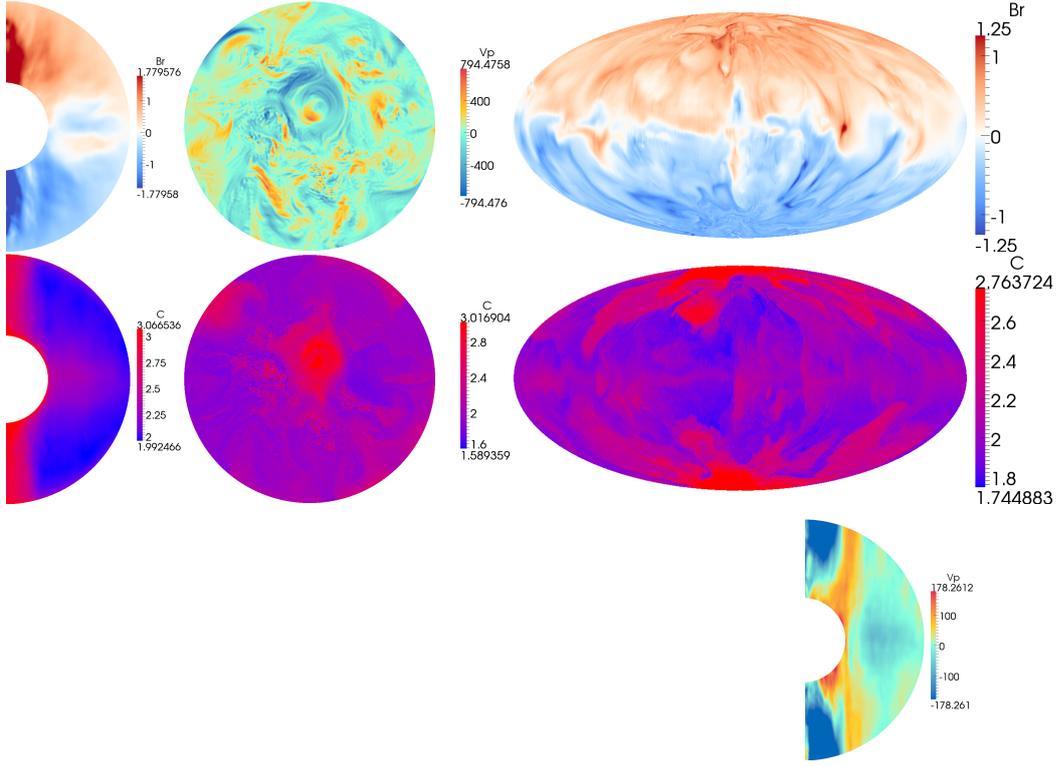

Figure 2: Behaviour of Case S4. The left hand column shows meridional sections (quantities have been averaged in time and azimuth), the middle column shows views from above of the plane z=$r_i$+0.5, and the third column shows the surface of the sphere in a Molleweide projection (all these are at $t = 0.0712324$). The top row shows radial magnetic field $B_r$, azimuthal velocity $v_\phi$ and radial field $B_r$ at $r_o$. Second row shows temperature with red representing higher temperature. The last figure shows azimuthal velocity $v_\phi$ averaged in azimuth and time.

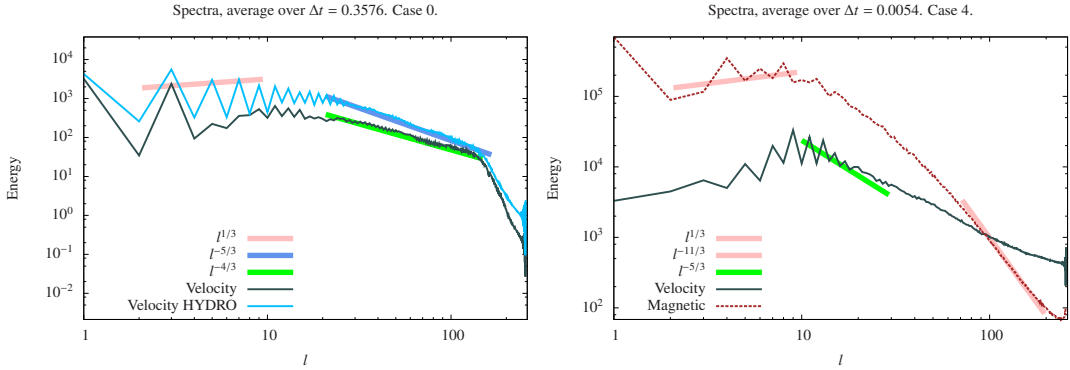

Figure 3: (a) Kinetic energy spectra (volume averaged) as a function of spherical harmonic degree $l$ are compared for Case S0 (a dynamo) and Case Hydro0: we see that the Lorentz force modifies the velocity field at all scales from $l = 1$ to $l \sim 150$. (b) Magnetic and kinetic energy spectra as a function of spherical harmonic degree $l$ for Case S4. The Ekman number is $3 \times 10^{-7}$ and $Pr_m = 0.05$. There is scale separation between the large scale magnetic field peaking at $l = 1$ and the velocity field with maximum energy at $l = 9$. Note that the magnetic energy, which is ten times larger in total than the kinetic energy, exceeds the latter for all $l < 100$.

## Force balance

The Coriolis and Lorentz forces in our most extreme model, Case S4, are shown in Figure4(a-c). Since convection is well-developed within the tangent cylinder, the forces are strongest here. In the figures we choose to examine the azimuthal average of the azimuthal component of all forces, which immediately removes the contribution from the pressure. We find evidence of a primary local balance between Coriolis and Lorentz forces, essentially the magnetostrophic or MAC force balance foreseen by Taylor (*8*), rarely seen in numerical simulations at more modest values of the control parameters. The use of a particularly small Ekman number means that the viscous forces play an insignificant role in the bulk of our dynamo, as expected for the Earth's core. While we see the largest scales are almost identical in Figures4a and Figure4b, there is a slight mismatch at the very smallest scales. To see this more clearly we examine the same quantity in the spectral domain, and examine the contributions by spherical harmonic degree. Figure4c shows a remarkable zeroth order balance between Coriolis and Lorentz forces at large scales, but at small scales that balance has been broken. We believe this effect has never been seen before, and arises when the diffusivities of magnetic field and momentum differ so much, as in our model. The Coriolis force is small scale, as it depends on **u**, whereas the Lorentz force is larger scale, depending quadratically on **B**. Although large scales can be balanced, the scale separation leads to a different balance at small scales where the role of nonlinear advection (Reynolds stresses) and accelerations are more pronounced. Even at the smallest scales the viscous force remains one order of magnitude smaller (in rms) than the Coriolis force. We believe that this behaviour represents well the appropriate dynamics of the core.

## Dissipation

All of our simulations have magnetic energy exceeding kinetic energy as is expected in the Earth's core, see Table 2. Of particular interest is the way that energy is dissipated. Ohmic dissipation dominates our simulations as the mechanism by which energy is returned to heat, with the exception of Case S0 where viscous and Ohmic dissipations are close to being equal. In Case S4 the effect of the well-developed convection in the tangent cylinder leads to a re-markable concentration of the Ohmic dissipation in this area (see Supplementary materials, Figure (S1)). Analysis of the distribution of dissipation shows that it is relatively constant in radius. When we analyse all our models together we observe a very precise power law scaling of magnetic dissipation with magnetic energy that appears to remain true over 3 decades of magnetic energies (Figure 3a). The favoured scaling fit to these non-dimensional data are of the form $D_{mag} = E_{mag}^{3/2}$ (with prefactor unity). We compare this law to the large library of runs supplied by Uli Christensen in Figure 3b. One can see that there is increased consistency of the fit to the scaling law in our low-$Pr_m$ runs, and that they agree with the data in Figure 3b only for high ohmic dissipations. Thus we have a considerable tightening of the systematic behaviour in these more realistic dynamos. The power law is consistent with the diffusionless theory of (*4, 9*) (see Supplementary Materials). We have a reasonably good estimate of the magnetic energy in the core as a result of Gillet et al's (*10, 11*) lower bound on field strength of 3mT. These values lead to an estimate of the magnetic energy in the core of at least $3 \times 10^{20}$J (This value is the value adopted by (*12*)). Using our scaling law (see Supplementary material) we find a probable value of 10TW, in line with the estimate given in (*9*) of 1-15TW. Such large values of dissipation suggest that the core is losing heat rapidly and suggest an inner core that is very young, likely less than 1Gyr in age. Dynamo action prior to this time is then challenging, though the new suggestion of magnesium dissolution in the early Earth (*13–15*) presents a possible resolution for the driving of the dynamo for the last 4.5Gyr.

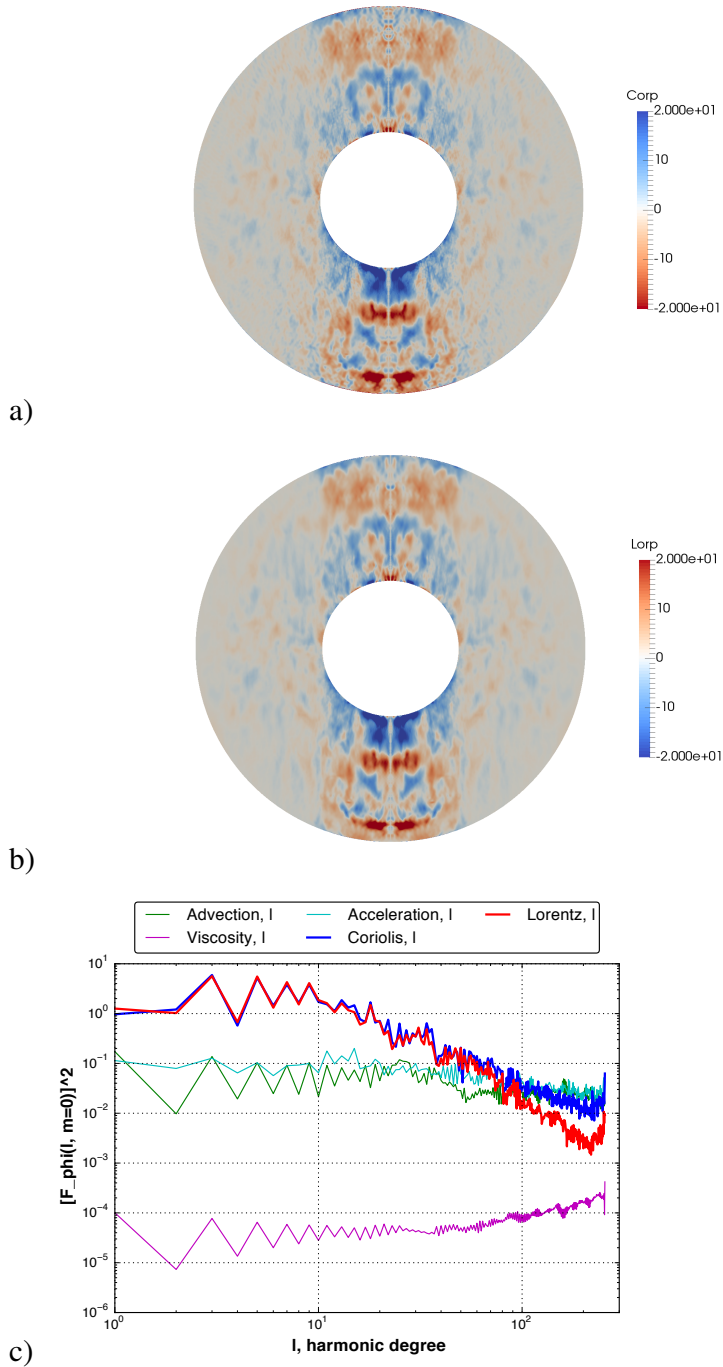

a)

b)

c)

Figure 4: Instantaneous force balances in model S4 at $t = 0.174$. a) and b) show the azimuthally averaged $\phi$ component of the (negative of the) Coriolis and Lorentz forces respectively, where the sign has been switched on the Coriolis force to aid comparison. For this average the pressure grandient vanishes identically and thus one can see almost perfect balance between the forces. c) Shows the same azimuthal component of all forces occurring in the Navier-Stokes equation when averaged over azimuth, squared, and integrated in non-dimensional radius (omitting 10% in radius at each boundary) as a function of spherical harmonic degree. This again removes the pressure component. The choice of radii is such that it excludes the boundary layers where there is a viscous-Coriolis-Lorentz balance. At the largest scales the balance between Coriolis and Lorentz forces is almost perfect. The rms viscous force is about 3 orders of magnitude smaller at the largest scales.

|    | $E_{mag}$ | $E_{kin}$ | $E_{mag}/E_{kin}$ | $D_{mag}$ | $D_{kin}$ | $D_{mag}/D_{kin}$ |
|----|-----------|-----------|-------------------|-----------|-----------|-------------------|
| S0 | 8.71e+04  | 2.91e+04  | 2.99              | 3.41e+07  | 4.04e+07  | 0.84              |
| S1 | 7.50e+05  | 2.38e+08  | 3.15              | 8.66e+08  | 5.04e+08  | 1.72              |
| S2 | 1.22e+07  | 6.35e+06  | 1.93              | 6.30e+10  | 1.36e+10  | 4.65              |
| S4 | 4.46e+06  | 5.51e+05  | 8.10              | 6.31e+09  | 6.18e+08  | 10.21             |

Table 2: Diagnostics measured in the simulations. $E_{mag}$ and $E_{kin}$ are the magnetic and kinetic energies, while $D_{mag}$ and $D_{kin}$ are the ohmic and viscous dissipations.

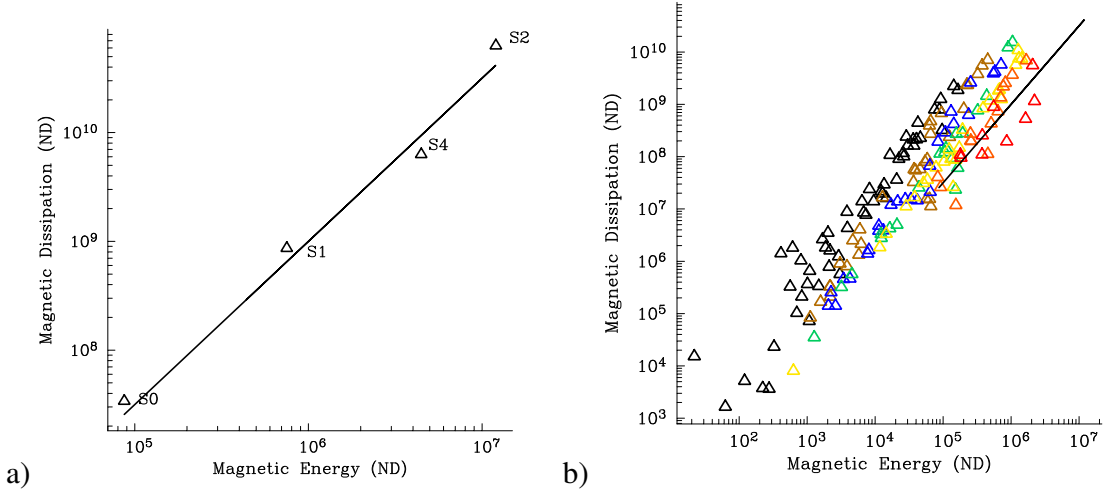

a)　　　　　　　　　　　　　　b)

Figure 5: Variation of magnetic dissipation $D_{mag}$ with magnetic energy $E_{mag}$ (a) for the new simulations reported herein (see Table 1) and (b) for the dataset from Uli Christensen. In (b) colours indicate the fraction of energy dissipated ohmically. The brightening spectrum starts with models (black) with predominantly viscous dissipation (ohmic dissipation $\leq 30\%$) and proceeds black-brown-blue-green-yellow-orange-red to the models dissipating mostly (red: 80-90%) ohmically. The straight line is a law of the form $D_{mag} = E_{mag}^{3/2}$.

20. **Acknowledgements**

    This study was supported by ERC Grant No. 247303 (MFECE) at ETH Zurich. Simulations were run on the Swiss National Supercomputing Center (CSCS) under the projects s225 and s369. The authors are grateful to Uli Christensen for sharing his database of dynamo runs.

# Supplementary Material

## Methods

### Governing equations and non-dimensionalization

We adopt the Boussinesq approximation for convection-driven, rotating magnetohydrodynamics, which results in the following non-dimensional equations:

$$
\left\{
\begin{array}{rcl}
\left( Ro\, \dfrac{\partial}{\partial t} - E\boldsymbol{\nabla}^2 \right) \mathbf{u} & = & \mathbf{N}_u - \boldsymbol{\nabla}\hat{P}, \\[2mm]
\left( \dfrac{\partial}{\partial t} - \boldsymbol{\nabla}^2 \right) \mathbf{B} & = & \boldsymbol{\nabla} \times (\mathbf{u} \times \mathbf{B}), \\[2mm]
\left( \dfrac{\partial}{\partial t} - q\boldsymbol{\nabla}^2 \right) T & = & \varepsilon - \mathbf{u} \cdot \boldsymbol{\nabla} T,
\end{array}
\right.
\tag{1}
$$

where

$$
\mathbf{N}_u = Ro\, \mathbf{u} \times (\boldsymbol{\nabla} \times \mathbf{u}) + (\boldsymbol{\nabla} \times \mathbf{B}) \times \mathbf{B} + q\, Ra\, T\, \mathbf{r} - \hat{\mathbf{z}} \times \mathbf{u}.
$$

Variables $\mathbf{u}$, $\mathbf{B}$, $T$ are the velocity, magnetic field and temperature. $\boldsymbol{\nabla}\hat{P}$ is the modified pressure that contains information about conservative forces. The axis of rotation of the system is $z$ and $\hat{\mathbf{z}}$ is a unit vector in its direction. Time is denoted as $t$. A uniform heat source $\varepsilon$ is included. Incompressibility conditions $\boldsymbol{\nabla} \cdot \mathbf{B} = 0$ and $\boldsymbol{\nabla} \cdot \mathbf{u} = 0$ are integrated into the solution technique through use of a poloidal-toroidal decomposition of the vector field. Non-dimensional parameters are defined as:

$$
\begin{array}{lll}
\text{Magnetic Rossby number} & Ro = \eta/(2\Omega d^2), \\
\text{Ekman number} & E = \nu/(2\Omega d^2), \\
\text{Modified Rayleigh number} & Ra = g\,\alpha\,\Delta T\, d/(2\Omega\kappa), \\
\text{Roberts number} & q = \kappa/\eta.
\end{array}
\tag{2}
$$

The units of length, time, magnetic field and temperature for the non-dimensional governing equations are chosen as follows:

$$
r \to d\, r,\; t \to d^2/\eta\, t,\; B \to (2\Omega\rho_0\mu_0\eta)^{\frac{1}{2}}\, B,\; T \to \Delta T\, T,\;\; d = r_o - r_i.
\tag{3}
$$

The following symbols denote the parameters of the system: $\boldsymbol{\Omega} = \Omega\hat{\mathbf{z}}$ is the rotation rate, $\mu_0$ is the permeability of free space, $\rho_0$ is the density, $\Delta T$ is the unit of temperature, $\nu$, $\kappa$ and $\eta$ are the kinematic viscosity, thermal diffusivity and magnetic diffusivity respectively, and $\alpha$ is the thermal expansivity. Gravity is assumed to vary linearly with radius and has value g on the outer boundary. The spherical coordinates are denoted $(r, \theta, \varphi)$.

**Boundary conditions and internal heating**

The modelled fluid is enclosed in a rotating spherical shell between radii $r_i$ and $r_o$ with $c = r_i/r_o = 0.35$. Both boundaries are no-slip and impermeable. The outer boundary is electrically insulating, the inner core has the same electrical conductivity as the outer core. In case S6_InsIC the inner core is insulating. The inner core temperature is kept constant at $T = 5.434$, the gradient of temperature on the outer core equals to $-2/(1 - c)$. A uniform heat source with $\varepsilon = 3q$ is adopted throughout the outer core.

**Diagnostics**

The kinetic and magnetic energies are defined as $E_{kin} = \frac{1}{2} \int u^2 \mathrm{d}V$ and $E_{mag} = \frac{1}{2Ro} \int B^2 \mathrm{d}V$, where the volume integral is over the entire outer core. The viscous and ohmic dissipations are defined as $D_{kin} = \frac{E}{Ro} \int (\nabla \wedge \mathbf{u})^2 \mathrm{d}V$ and $D_{mag} = \frac{1}{Ro} \int (\nabla \wedge \mathbf{B})^2 \mathrm{d}V$. The magnetic Reynolds number $R_m$ is $ud/\eta$.

**Numerical setup**

We solve the governing equations using a parametrization in spherical harmonics up to degree and order 255 for the angular component and 528 finite difference points in radius. A second order predictor-corrector scheme is used for the time integration (*16*). The timestep is adaptive and varies throughout the run. Parallelisation is carried out in radius. In the linear parts of the code, data is split over the spherical harmonics. 528 cores were used simultaneously for one simulation. The bulk of the simulations and visualizations were performed on the supercomputer Piz Daint (Cray XC 30) at Swiss National Supercomputing Center. The code was originally developed by Willis (*17*) and then subsequently optimized for the Cray XC 30 and successfully benchmarked against other dynamo codes (*18*).

## Dissipation in the Earth

Based on our non-dimensionalisation the relationship between dimensionless energies and true energies is the following (starred quantities are dimensional and unstarred are nondimensional): The kinetic energy is

$$E_{kin}^* = \int_V \rho(u^*)^2/2dV^* = L\rho\eta^2 \int u^2/2dV = L\rho\eta^2 E_{kin} \tag{4}$$

The magnetic energy is

$$E_{mag}^* = \int_V \frac{(B^*)^2}{(2\mu_0)}dV^* = L\rho\eta^2 \frac{1}{Ro} \int \frac{B^2}{2}dV = L\rho\eta^2 E_{mag} \tag{5}$$

The magnetic dissipation is

$$D_{mag}^* = \int (J^*)^2/\sigma dV^* = \eta \int (\nabla^* \wedge B^*)^2/\mu_0 dV^* = \eta^3\rho/L \frac{1}{Ro} \int (\nabla \wedge B)^2 dV = (\eta^3\rho/L)D_{mag} \tag{6}$$

Thus if we use the lower bound for the magnetic energy of $3 \times 10^{20}$J together with the rule from Figure (5) we have

$$\eta^{-3}\rho^{-1}LD_{mag}^* = (\rho^{-1}\eta^{-2}/LE_{mag}^*)^{3/2} \tag{7}$$

$$D_{mag}^* = 16 \quad \rho^{-1/2}L^{-5/2}(E_{mag}^*)^{3/2} \tag{8}$$

$$\sim 10^{13}W \tag{9}$$

## The dissipation-energy relation

Here we remark that the scaling relationship $D_{mag} \sim E_{mag}^{3/2}$ can be related to the magnetic field scaling laws suggested by (4, 9). In (9) the ohmic dissipation is related to the energy flux $Fq_0$ by

$$D = f_{\text{ohm}}Fq_0. \tag{10}$$

The Lorentz number scaling of **B** with $Fq_0$ is very close to $1/3$ and so magnetic energy will scale with a $2/3$ exponent. Using (10) with $f_{\text{ohm}} = 1$ gives $E_{mag} \sim D_{mag}^{2/3}$, leading to the required relationship. We note that stress-free calculations of (19) give a B-scaling exponent of 0.37. One can see that a relationship $D_{mag} \sim E_{mag}^{4/3}$ would lead to a scaling exponent of $3/8 = 0.375$ which is essentially the same. However this law is not really preferred by our data, instead we find more accord with the original exponent of (*4*). Thirty-eight out of forty of the models presented in (*19*) have $D_{kin} > D_{mag}$ and thus do not dissipate primarily through Ohmic dissipation. This may contribute to the discrepancy.

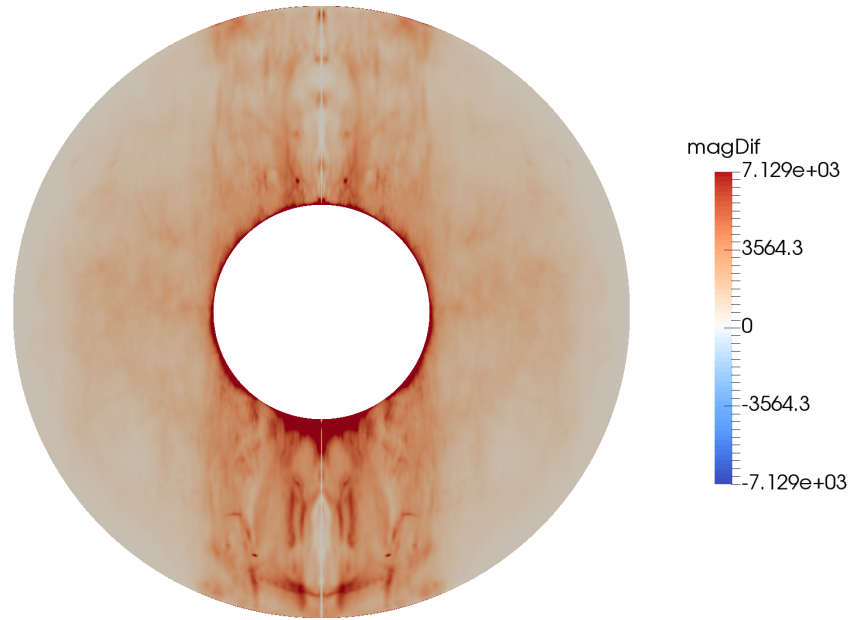

Figure S1: Azimuthally averaged Ohmic dissipation for model S4. The importance of the tangent cylinder is apparent.